%% file: main.tex
\documentclass[10pt,twocolumn,letterpaper]{article}

\usepackage{wacv}              

\usepackage{graphicx}
\usepackage{orcidlink}
\usepackage{amsmath}
\usepackage{amssymb}
\usepackage{booktabs}
\usepackage{adjustbox}
\usepackage{arydshln}
\usepackage[accsupp]{axessibility} 

\usepackage[capitalize]{cleveref}
\crefname{section}{Sec.}{Secs.}
\Crefname{section}{Section}{Sections}
\Crefname{table}{Table}{Tables}
\crefname{table}{Tab.}{Tabs.}

\def\wacvPaperID{374} 
\def\confName{WACV}
\def\confYear{2025}

\input{figures.tex}

\newcommand{\x}{\mathbf{x}}
\newcommand{\z}{\mathbf{z}}
\newcommand{\s}{\mathbf{s}}

\newcommand{\qzx}{q_\mathbf{\phi}(\z|\x)}
\newcommand{\pzx}{p_\mathbf{\theta}(\z|\x)}
\newcommand{\pxz}{p_\mathbf{\theta}(\x|\z)}

\newcommand{\pz}{p_\mathbf{\theta}(\z)}

\newcommand{\hdn}{HDN\textsubscript{36}}

\begin{document}

\title{Unsupervised Denoising for Signal-Dependent \\and Row-Correlated Imaging Noise} 

\author{Benjamin Salmon
\orcidlink{0000-0002-5919-0158}
and Alexander Krull
\orcidlink{0000-0002-7778-7169}
\\
School of Computer Science, University of Birmingham\\
{\tt\small brs209@student.bham.ac.uk, a.f.f.krull@bham.ac.uk}\\
\small\url{https://github.com/krulllab/COSDD}
}



\maketitle
\input{sections/abstract}
\input{sections/intro}
\input{sections/related_work}
\input{sections/background}
\input{sections/method}
\input{sections/experiments}
\input{sections/conclusion}

{\small
\bibliographystyle{ieee_fullname}
\bibliography{main}
}

\end{document}


\title{Unsupervised Denoising for Signal-Dependent \\and Row-Correlated Imaging Noise \\ Supplementary Material} 

\author{Benjamin Salmon and Alexander Krull\\
School of Computer Science, University of Birmingham\\
{\tt\small brs209@student.bham.ac.uk, a.f.f.krull@bham.ac.uk}\\
\small\url{https://github.com/krulllab/COSDD}
}



\maketitle

\input{supp_sections/Latent_variables_represent_clean_images}\input{supp_sections/Training_and_inference}\input{supp_sections/Architecture}\input{supp_sections/Baselines}
\input{supp_sections/Simulated_data}
\input{supp_sections/qualitative_results}

{\small
\bibliographystyle{ieee_fullname}
\bibliography{supp}
}

%% file: figures.tex
\newcommand\figNoiseMatrix{
\begin{figure}[h]
    \centering
    \includegraphics[width=1\linewidth]{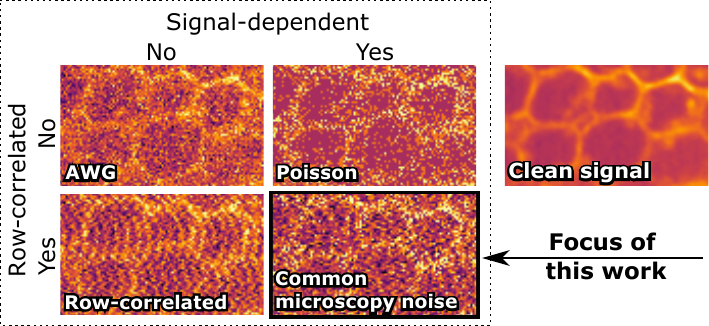}
    \caption{
    Noise is often assumed to be spatially uncorrelated and/or signal-independent such as additive white gaussian noise (AWG).
    Such assumptions greatly facilitate self/unsupervised denoising.
    Unfortunately, real noise in microscopy is usually neither.
    We present the first unsupervised denoising approach for this challenging scenario.
    }
    \label{fig:noisematrix}
    \vspace{-3mm}
\end{figure}
}

\newcommand\figExplainer{
\begin{figure}[h]
    \centering
    \includegraphics[width=1\linewidth]{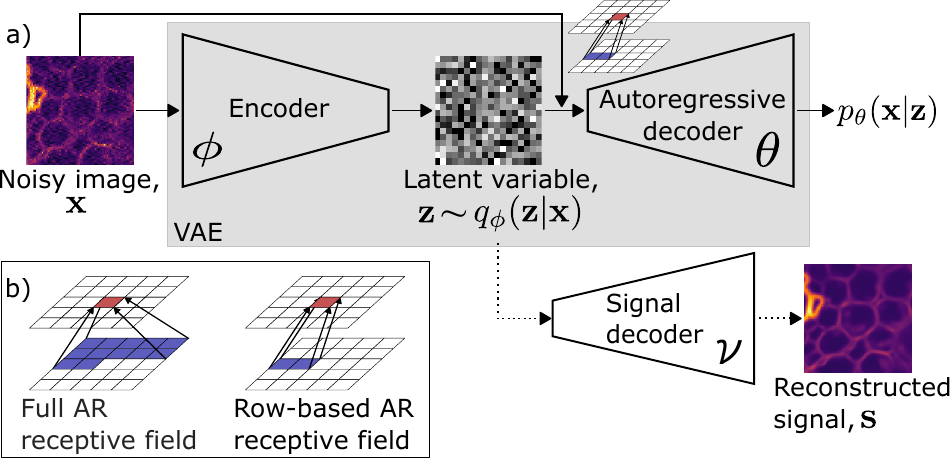}
    \caption{
    \label{fig:explainer}
    {\bf a):}
    A Variational Autoencoder~\cite{chenvariational} (solid arrows) is trained to model the distribution of noisy images $\mathbf{x}$. 
    The autoregressive (AR) decoder models the noise component of the images, while the latent variable models only the clean signal component $\mathbf{s}$.
    In a second step (dashed arrows), our novel \emph{signal decoder} is trained to map latent variables into image space, producing an estimate of the signal underlying $\mathbf{x}$.
    {\bf b):}
    To ensure that the decoder models only the imaging noise and the latent variables capture only the signal, we modify the AR decoder's receptive field.
    In a full AR receptive field (\cref{eq:full-auto}), each output pixel (red) is a function of all input pixels located above and to the left (blue). In our decoder's row-based AR receptive field (\cref{eq:row_decoder}), each output pixel is a function of input pixels located in the same row, which corresponds to the row-correlated structure of imaging noise.
    }
\end{figure}
}

\newcommand\tabpsnr{
\begin{table*}[h]
\centering
\caption{
Comparison to baseline methods using Peak Signal-to-Noise Ratio (PSNR), higher is better. The best results for methods that do not require paired images are printed in \textbf{bold}, and best results overall are \underline{underlined}.
HDN is a variant of \hdn~that is intended for spatially uncorrelated noise, but was applied to two datasets with spatially correlated noise by the original authors.
These extra results are presented in brackets.
Note that the \emph{Mouse Nuclei} dataset contains spatially uncorrelated noise and that \hdn~failed to train on the \emph{Mouse Actin} dataset.
CARE and N2N are supervised denoisers requiring paired images, unlike the other baselines.
}
\begin{adjustbox}{width=1\textwidth}
\begin{tabular}{l|ccc:ccc:cc}
\multicolumn{1}{l|}{} & \multicolumn{3}{c:}{EMCCD} & \multicolumn{3}{c:}{LSCM} & \multicolumn{2}{c}{Simulated} \\
\multicolumn{1}{l|}{} & Conv. A~\cite{broaddus2020removing} & Conv. B~\cite{prakash2020pn2v} & Mouse Actin~\cite{prakash2020pn2v} & Mouse Nuclei*~\cite{prakash2020pn2v} & Actin Conf.~\cite{hagen2021fluorescence} & Mito Conf. ~\cite{hagen2021fluorescence} & FFHQ Stripe~\cite{karras2019style}  & FFHQ Checkerb.~\cite{karras2019style} \\ \hline
AP-BSN~\cite{lee2022ap} & 22.30 & 24.10 & 29.55 & 35.41 & 26.89 & 24.94 & 28.22 & 13.19 \\
SN2V~\cite{broaddus2020removing} & 30.29 & 31.67 & 32.80 & 36.62 & 23.30 & 26.41 & 29.52 & 19.11 \\
N2V2~\cite{hock2022n2v2} & 29.36 & 36.72 & 34.23 & 35.88 & 26.71 & 25.89 & 33.08 & 29.22 \\
HDN$_{36}$ (HDN)~\cite{prakashinterpretable} & 31.41 & 37.21 (37.39) & - (34.12) & (36.87) & 26.84 & 26.51 & 32.64 & 29.61 \\
HDN$_{36}$ \textit{large} (HDN \textit{large})~\cite{prakashinterpretable}  & 31.80 & 37.92 & 34.14 & (38.12) & 27.17 & 26.15 & 34.54 & 25.51 \\
Ours \textit{small} & 34.42 & 38.99 & 36.50 & 39.56 & 27.35 & 27.49 & 32.87 & 33.51 \\
Ours \textit{large} & \underline{\textbf{37.49}} & \underline{\textbf{44.10}} & \underline{\textbf{39.23}} & \underline{\textbf{42.98}} & \textbf{27.41} & \textbf{27.50} & \textbf{35.66} & \textbf{36.27} \\ 
\hdashline
CARE~\cite{weigert2018content} & 31.56 & 36.71 & 34.20 & 36.58 & \underline{29.44} & \underline{27.55} & \underline{36.46} & \underline{36.89}  \\
N2N~\cite{lehtinen2018noise2noise} & 28.22 & 36.85 & 34.60 & 37.33 & - & - & 36.18 & 36.43
\end{tabular}
\end{adjustbox}
\label{tabpsnr}
\end{table*}
}

\newcommand\figVisResults{
\begin{figure*}[h]
    \centering
    \includegraphics[width=0.9\linewidth]{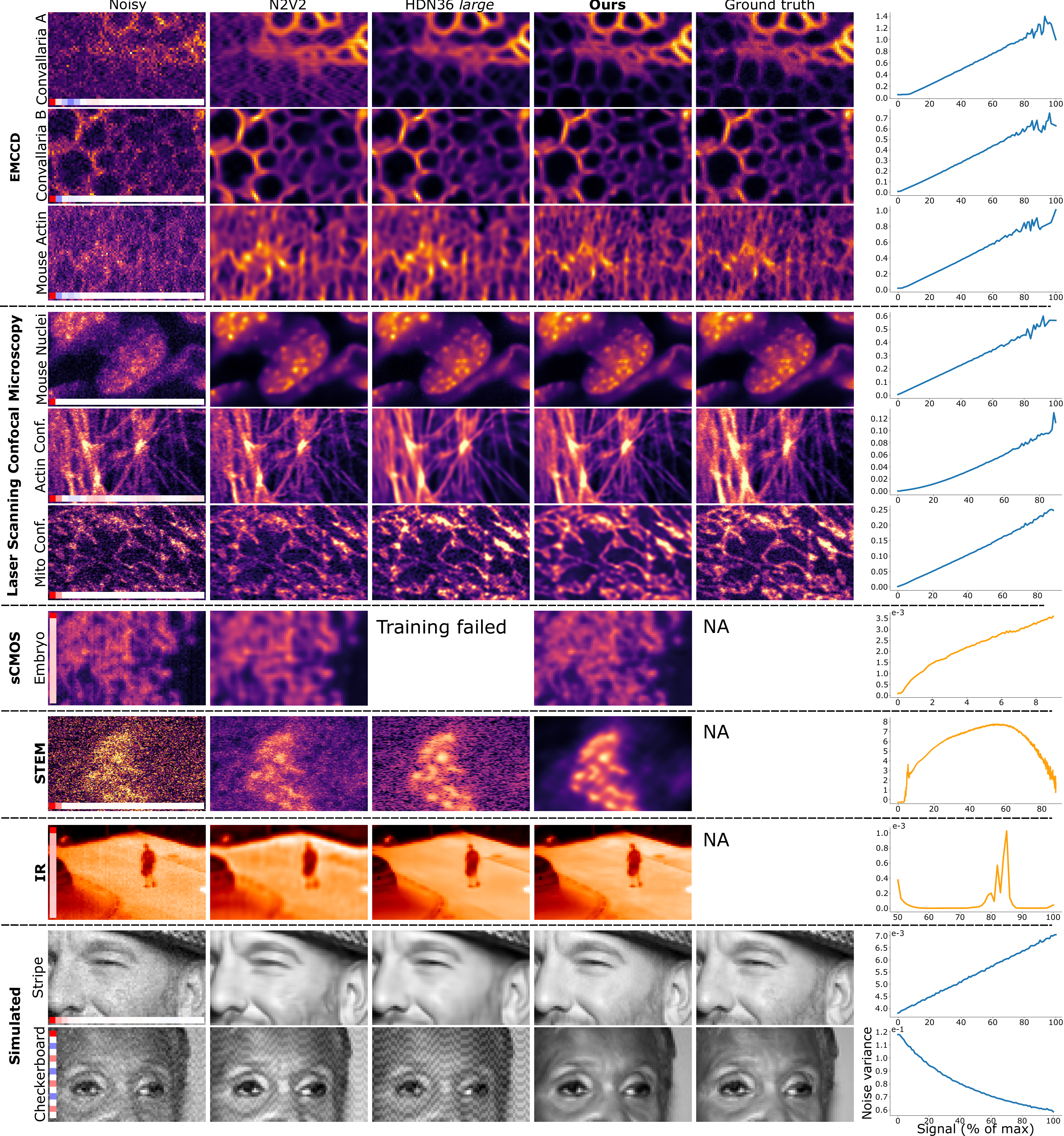}
    \caption{Visual results from our method and two unpaired baselines on all datasets. 
    The spatial autocorrelation of the noise is overlaid on each noisy image, with red indicating a positive correlation and blue indicating a negative correlation.
    The direction of the correlation is given by the orientation of the autocorrelation bar.
    Additionally, the signal dependence of the noise in each dataset is shown in the right-hand column.
    The horizontal axis of these graphs is the clean signal intensity as a percentage of the maximum, while the vertical axis is the variance of noisy pixel values recorded for these signal intensities.
    Ground truth must be used to calculate signal dependence, so orange lines are used where denoised images from our method are used as pseudo-ground truth.}
    \label{fig:visResults}
\end{figure*}
}

\newcommand\figRFsize{
\begin{figure}[h]
    
    \centering
    \includegraphics[width=1\linewidth]{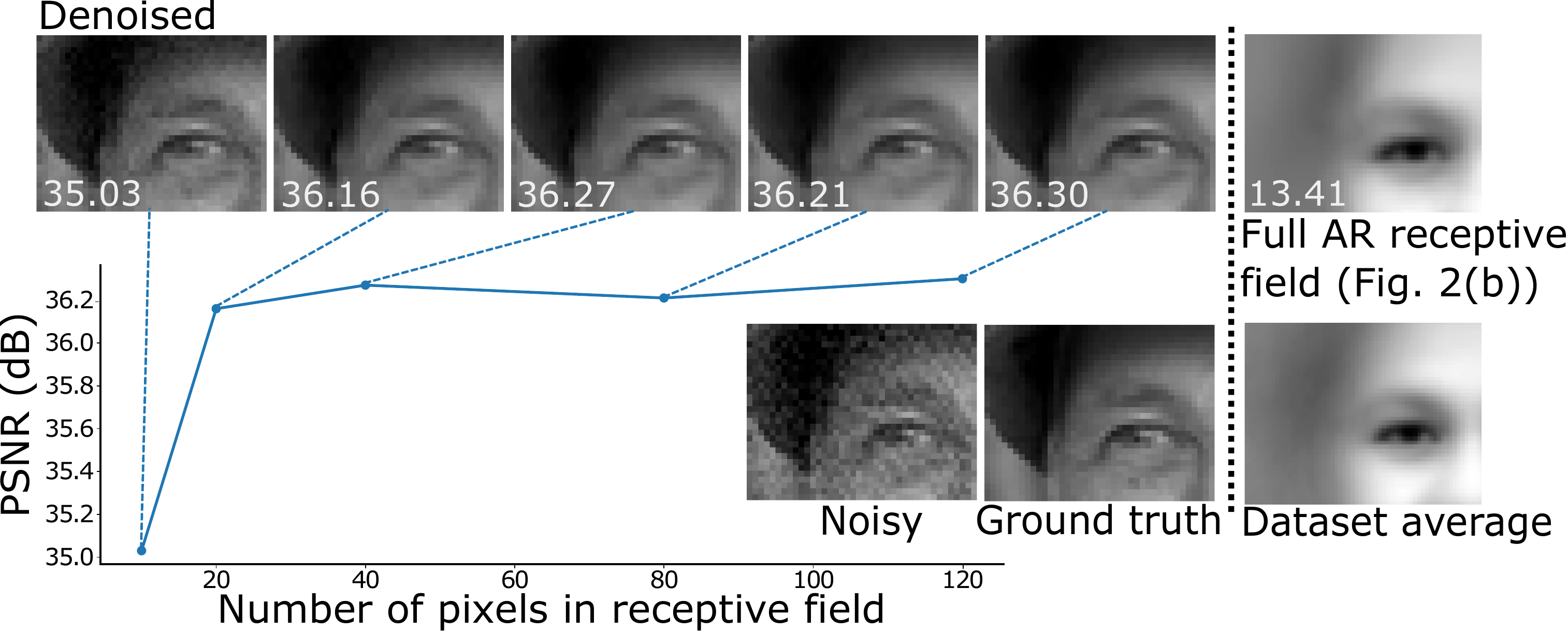}
    \caption{
    The \emph{FFHQ - Checkerboard} dataset wass denoised 5 times by varying the number of pixels covered by the AR decoder's receptive field.
    The images show denoising results for each receptive field size with PSNR overlaid.
    Additionally, an image denoised using a full AR receptive field is included on the right.
    In this situation, the signal decoder is given completely uninformative inputs and learns to output the mean of the training dataset.
    }
    \label{fig:rfSize}
    \vspace{-3mm}
\end{figure}
}

\newcommand\figNoiseRecon{
\begin{figure}[h]
\vspace{-2mm}
    \centering
    \includegraphics[width=1\linewidth]{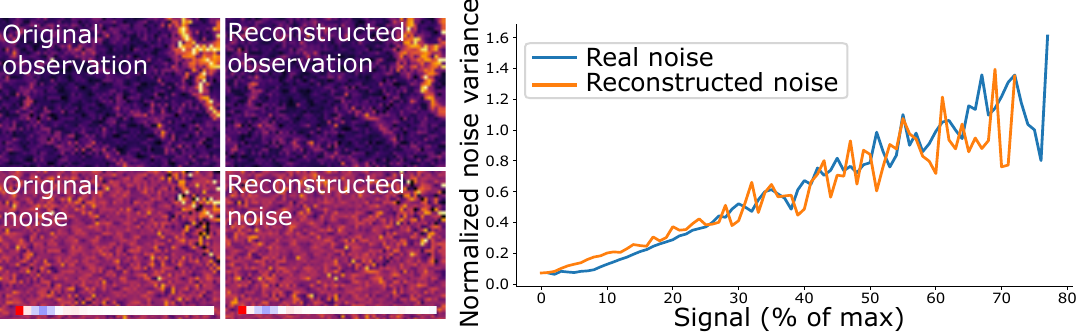}
    \caption{
    A noisy image from the \emph{Convallaria A} dataset was encoded and decoded to produce a reconstructed observation and an artificial noise sample.
    The sampled noise has spatial autocorrelation and signal dependence that match those of the real noise, indicating that the AR decoder has learnt an accurate noise model.
    \vspace{-6mm}
    }
    \label{fig:noiserecon}
\end{figure}
}

%% file: sections/abstract.tex
\begin{abstract}
Accurate analysis of microscopy images is hindered by the presence of noise.
This noise is usually signal-dependent and often additionally correlated along rows or columns of pixels.
Current self- and unsupervised denoisers can address signal-dependent noise, but none can reliably remove noise that is also row- or column-correlated.
Here, we present the first fully unsupervised deep learning-based denoiser capable of handling imaging noise that is row-correlated as well as signal-dependent.
Our approach uses a Variational Autoencoder (VAE) with a specially designed autoregressive decoder.
This decoder is capable of modeling row-correlated and signal-dependent noise but is incapable of independently modeling underlying clean signal.
The VAE therefore produces latent variables containing only clean signal information, and these are mapped back into image space using a proposed second decoder network.
Our method does not require a pre-trained noise model and can be trained from scratch using unpaired noisy data.
We benchmark our approach on microscopy datatsets from a range of imaging modalities and sensor types, each with row- or column-correlated, signal-dependent noise, and show that it outperforms existing self- and unsupervised denoisers.
\vspace{-2mm}
\figNoiseMatrix
\end{abstract}

%% file: sections/intro.tex
\section{Introduction}
\label{sec:intro}

Microscopy has wide ranging applications in natural sciences.
As imaging technology is routinely pushed to its limits, microscopy images become affected by the presence of noise.
This poses a challenge in downstream analysis, making denoising an important preprocessing step.
Over the years, various denoising approaches have been devised~\cite{laine2021imaging}, from traditional filter-based methods~\cite{dabov2006image} to supervised deep learning-based methods~\cite{weigert2018content}.

Supervised deep learning-based methods excel with respect to the quality of their output, however, they require paired training data, typically consisting of pairs of noisy and clean images.
Unlike in applications with consumer images (photographs), such paired data is often unobtainable in microscopy~\cite{izadi2023image, song2021noise2void, xu2021deformed2self}.
Consequently, the applicability of supervised denoising methods in many areas of scientific imaging is limited.

Self- and unsupervised methods (\eg~\cite{batson2019noise2self,krull2019noise2void,prakash2020fully,prakashinterpretable,eom2023statistically}) have been proposed as a solution to the lack of paired training data. 
They can be trained directly on the data that is to be denoised and have substantially improved the practicality of denoising in scientific imaging.
These methods separate the imaging noise from the underlying signal by making assumptions about the statistical nature of the noise.
Typically, they assume the noise is $(i)$ signal-independent (purely additive and occurs separate from the underlying signal)~\cite{salmon2022towards}, or $(ii)$ spatially uncorrelated (unstructured and occurs separately for each pixel)~\cite{krull2019noise2void, krull2020probabilistic, prakash2020fully, huang2021neighbor2neighbor, quan2020self2self}. 
Examples of $(i)$ and $(ii)$ are in \cref{fig:noisematrix}.

In practice, $(i)$ is often broken by the presence of Poisson shot noise~\cite{laine2021imaging}.
Moreover, many popular scientific cameras and imaging setups break $(ii)$ by producing row- or column-correlated noise (see \cref{sec:noise} \textbf{Row-Correlated Noise}).
These types of noise cannot be removed with basic self- or unsupervised methods.
Recently, variants of these methods for spatially correlated noise have been proposed, but they are limited to locally correlated noise and can come at the expense of reduced reconstruction quality~\cite{broaddus2020removing,prakashinterpretable}.

In this paper, we present the first unsupervised deep learning-based denoiser capable of reliably removing signal-dependent noise that is correlated along rows or columns of pixels, as it commonly occurs in microscopy data.
The approach is illustrated in \cref{fig:explainer}.
Our method requires neither examples of noise-free images, which can be impossible to obtain (\eg~\cite{wang2020noise2atom}), nor pre-trained noise models~\cite{prakash2020fully, salmon2022towards, krull2020probabilistic} or hand-crafted priors~\cite{wang2020noise2atom, dabov2006image, gu2014weighted}.
Furthermore, we do not rely on blind-spot approaches~\cite{broaddus2020removing} or subsampling~\cite{huang2021neighbor2neighbor, lee2022ap} techniques that degrade image quality and limit denoising performance.


Our method is based on the representation learning technique proposed by Chen~\etal~\cite{chenvariational}, who revealed how a Variational Autoencoder (VAE)~\cite{kingma2014auto} prefers to avoid using latent variables
to describe structures that could instead be modeled locally by its decoder.
For an autoregressive (AR) decoder such as PixelCNN~\cite{van2016conditional, gulrajani2016pixelvae}, these are any inter-pixel dependencies that lie within its receptive field.
Consequently, latent variables will only represent information that the decoder cannot model, \ie, dependencies that span beyond its receptive field.

We take advantage of this behavior by designing a decoder that is capable of modeling exactly what we want to be excluded from our latent variables -- the imaging noise -- while being incapable of modeling the remainder of the data.
Specifically, we use an AR decoder that can only model axis-aligned structures because its receptive field spans only a row or column of the image.
This trains our model to exclude row- or column-correlated noise from the latent variables, while encouraging it to include the statistics of the underlying signal.
We then propose a second network, termed \emph{signal decoder}, that is trained to map these latent variables back into image space, thus producing denoised images.
Following Lehtinen~\etal~\cite{lehtinen2018noise2noise}, who showed that noisy images can be used as training targets for denoisers, we train the signal decoder alongside the VAE using the original noisy data as targets.

Using real microscopy data recorded with various imaging modalities, we demonstrate that our denoiser achieves state-of-the-art results compared to other unsupervised methods.
Furthermore, we show that the method is not sensitive to the length of the AR decoder's receptive field, as long as it is above a minimum, and that samples from the trained noise model have the same autocorrelation and signal-dependence as real noise.


%% file: sections/related_work.tex
\section{Related Work}
\label{sec:relwork}
In the following, we will focus on methods that do be trained without paired training data.

\noindent{\textbf{Self-Supervised Denoising}}\quad
Most self-supervised denoisers use the blind-spot approach~\cite{batson2019noise2self, eom2023statistically, krull2019noise2void, quan2020self2self, wang2022blind2unblind}, where a network learns a denoising function when trained to predict the value of a pixel from surrounding pixels.
Another technique trains the network to predict a subset of randomly sampled pixels from another~\cite{huang2021neighbor2neighbor}.
For photon-counting data, Krull~\etal~\cite{krull2024image} proposed removing photons and training a network to predict them with a Poisson distribution.
These techniques all exploit a property of spatially uncorrelated noise, which is that the noise in one pixel cannot be predicted from the noise in other pixels.

To address spatially correlated noise, Structured Noise2Void (SN2V)~\cite{broaddus2020removing} extended the blind-spot approach to also mask pixels containing noise that is correlated with the noise in the pixel being predicted.
Lee~\etal~\cite{lee2022ap} also extended the blind-spot approach by subsampling pixels to break up noise structures, making noise effectively spatially uncorrelated and ready for traditional blind-spot denoising.
N2V2~\cite{hock2022n2v2} constrained a blind-spot network's ability to reconstruct high-frequency content.

There are also methods that add two independent noise samples to already-noisy images, training networks to denoise by mapping between these doubly-noisy pairs~\cite{moran2020noisier2noise,  pang2021recorrupted}. 
While theoretically applicable to signal dependent, spatially correlated noise, practical implementation has not been demonstrated.
The obstacle lies in obtaining a suitable noise model to sample from, particularly when signal dependent and spatially correlated noise must be described by a covariance matrix (\cite{pang2021recorrupted}).

\noindent{\textbf{Diffusion-based Denoising}}\quad
In recent years, diffusion models have been proposed as priors for solving inverse problems, such as denoising, in a Bayesian way~\cite{chungdiffusion, elad2023image,song2021solving}.
However, such methods require clean training data and a known mathematical forward model of the corruption process during inference, both of which are unavailable in many microscopy applications.
Very recently, methods have been proposed to remove at least the requirement for clean training data by assuming simple additive Gaussian noise~\cite{aali2023solving} or focusing on inpainting problems~\cite{daras2024ambient}.

\noindent{\textbf{GAN-based Denoising}}\quad
Another approach of denoising trains a conditional generative adversarial network~\cite{mirza2014conditional} to model the distribution of clean images given noisy images.
This can be done using a dataset of unpaired noisy and clean images, provided that the generator is forced to output an estimate of the clean signal underlying its noisy input, and not a random clean image.
Various methods have been proposed to ensure this, including minimizing the perceptual difference between the generator's input and output~\cite{yang2021unpaired}, or maximizing their mutual information~\cite{wang2022optimal}.
While these methods have more relaxed requirements than supervised denoisers, obtaining a dataset of clean images can be a challenge in microscopy, even if it is unpaired.
This is because the clean images must follow the same distribution as the signal content in the noisy training data.
That means,~\eg, the same cell type, imaging conditions, \etc.

\noindent{\textbf{VAE-based Denoising}}\quad
These methods train a latent variable model of noisy data using a VAE and a pre-trained explicit noise model.
Through the use of the noise model, the VAE is trained to represent clean signals with its latent variables.
In DivNoising~\cite{prakash2020fully}, the noise model 
assumes the noise is spatially uncorrelated, \ie, noise is generated in each pixel independently. 
Later, HDN~\cite{prakashinterpretable} was proposed as an extension to DivNoising that used a VAE with a hierarchy of latent variables~\cite{sonderby2016ladder}.
It was found that short-range spatially correlated noise structures were modeled by only the bottom levels of the hierarchy and could be removed by preventing these latent variables from using information from the encoded input.
This technique is known as \hdn.

Next, Autonoise~\cite{salmon2022towards} was proposed to remove spatially correlated but signal-independent noise by replacing the pixel-independent noise model in HDN with a CNN-based AR noise model~\cite{van2016conditional}.
This noise model was pre-trained using samples of pure noise which can be obtained by, \eg, imaging without light.
The method was therefore able to remove noise of any spatial correlation, but could not be applied to signal-dependent noise.



%% file: sections/background.tex
\section{Background}
Before we come to describe our method, we will first formalise the type of noise that we aim to remove, then give some background on VAEs for representation learning.

\subsection{Image Formation and Imaging Noise}
\label{sec:noise}
We can express image formation as a two step process.
The clean image, or as we also refer to it, the \emph{signal}, $\s=(s_{1,1}, \dots, s_{N,M})$, is drawn from a distribution $p(\s)$ and then subjected to noise, producing the noisy image $\x=(x_{1,1}, \dots, x_{N,M})$ as drawn from the noise distribution $p(\x|\s)$.
Here, $s_{i,j}$ and $x_{i,j}$ correspond to the respective pixel values at position $(i,j)$, and $N$ and $M$ are the number of rows and columns in the image, respectively.

Generally, we assume that the imaging noise is zero-centered, that is, the expected value of a noisy image equals the signal; $\mathbb{E}_{p(\x|\s)}[\x] = \s$. 
Note that zero-centered noise is a weak assumption that is widely shared in literature~(\eg~\cite{krull2019noise2void, laine2019high,eom2023statistically}).
Another way of looking at this assumption is that the sensor or camera is assumed to behave linearly so that an increase in $s_{i,j}$ (\eg by increasing the light intensity) will, on average, result in a proportional increase in $x_{i,j}$.

\noindent\textbf{Additive White Gaussian Noise}\quad
The most basic traditionally used noise model is additive white Gaussian (AWG) noise, see~\eg \cite{dabov2006image,buades2005non,metzler2018unsupervised}.
We can write the probability distribution for noisy observations in an AWG noise model as
\begin{equation}
    p(\x|\s) 
    = \prod_{i=1}^{N} \prod_{j=1}^{M} p(x_{i,j}|s_{i,j}),
    \label{eq:pixel-indep}
\end{equation}
formulated as a product over pixels.
In general, we refer to any type of noise model that can be factorised according to \cref{eq:pixel-indep} as spatially uncorrelated.
In AWG noise, $p(x_{i,j}|s_{i,j})$ corresponds to a normal distribution centered at $s_{i,j}$ with variance fixed for all $s_{i,j}$.
In general, we call noise with a variance that does not depend on the underlying signal as
signal-independent.
Noise for which this is not the case will be referred to as signal-dependent.

Denoising approaches have assumed noise to be both spatially uncorrelated and signal-independent.
Unfortunately, real imaging noise rarely follows these assumptions.

\noindent\textbf{Poisson Shot Noise}\quad
Most imaging systems are 
affected by Poisson shot noise~\cite{laine2021imaging}.
A Poisson noise model follows \cref{eq:pixel-indep}, with $p(x_{i,j}|s_{i,j})$ corresponding to a Poisson distribution with rate $\lambda=s_{i,j}$.
A popular variant on this model is the Poisson-Gaussian noise model (\eg~\cite{zhang2019poisson}), where $p(x_{i,j}|s_{i,j})$ is assumed to be a combination of Poisson shot noise and Gaussian read-out noise.

Poisson and Poisson-Gaussian noise models are spatially uncorrelated but signal-dependent, since the variance of the Poisson component in each pixel depends on the pixel's signal $s_{i,j}$.

\noindent\textbf{Row-Correlated Noise}\quad
While Poisson-Gaussian noise models are a popular choice in describing noise distributions, many imaging systems 
can produce noise that does not conform to \cref{eq:pixel-indep}, but is instead correlated along rows or columns of pixels.
For example, Scanning Transmission Electron Microscopy (STEM)~\cite{pennycook2011scanning} is prone to line artifacts caused by the slow reaction time of readout electronics~\cite{liao2006practical}, and Laser Scanning Confocal Microscopy (LSCM)~\cite{paddock2000principles} has been shown to produce artifacts in the scanning direction~\cite{herberich2012signal}.
Depending on their settings, Electron Multiplying Charge-Coupled Device (EMCCD)~\cite{denvir2003electron} cameras can produce horizontally correlated readout noise~\cite{broaddus2020removing}. 
The scientific Complementary Metal-Oxide-Semiconductor (sCMOS)~\cite{moomaw2013camera} cameras that are popular in optical microscopy~\cite{mandracchia2020fast} have separate amplifiers for each column of pixels, leading to correlated noise within each column~\cite{zhang2021characterizing}.
In addition to microscopy, the detectors used in infrared imaging systems, \eg microbolometers, also commonly use separate column amplifiers and suffer from similar noise structures~\cite{dupont2009fpn}. 
Examples of noisy images from these modalities along with plots of their spatial autocorrelation and signal-dependence can be found in \cref{fig:visResults}.

We propose to describe such noise as
\vspace{-1mm}
\begin{equation}
    p(\x|\s) 
    = \prod_{i=1}^{N} \prod_{j=1}^{M} 
    p(x_{i,j}|\s, x_{i,1},\dots,x_{i,j-1}),
    \label{eq:structured}
\end{equation}
where 
$p(x_{i,j}|\s, x_{i,1},\dots,x_{i,j-1})$
is the distribution of possible noisy pixel values $x_{i,j}$ conditioned on the signal, as well as all ``previous'' values in the same row, $i$.
While this model can describe interactions between pixels within the same row, pixels in different rows are (conditionally) independent (given a signal $\s$).
Note that in this formulation, a pixel $x_{i,j}$ can depend on not only the signal in pixel $(i,j)$, as with shot noise, but on the entire image for more complex interactions.
In this work, we address the type of noise described in~\cref{eq:structured}, which we believe is a good model for many real scientific imaging data.

\subsection{Latent Variable Models and VAEs}
\label{sec:vae}
A latent variable model, with parameters $\theta$, defines a probability distribution, $p_\theta(\x) \propto \pxz\pz$, over observed variables $\x$ via latent variables $\z$. 
This can be used to represent a data generation process in which a value $\z$ is first sampled from the prior $\pz$ and a value $\x$ is sampled from the conditional distribution $p_\theta(\x|\z)$.

A VAE can be used to simultaneously optimize the model's parameters and approximate the posterior, $p_\theta(\z|\x)$, via maximization of a lower bound on the marginal log-likelihood, 
\begin{equation}
\label{eq:VAEloss}
    \mathcal{L}(\theta,\phi)=\mathbb{E}_{\qzx}[\log \pxz)] - KL(\qzx)\Vert\pz),
\end{equation}
where the second term on the RHS is the Kullback-Leibler divergence~\cite{kullback1951information} from the prior to an approximate posterior $\qzx$.
In \cref{eq:VAEloss}, the first term is known as the reconstruction error, the second term as the regularization error, $\qzx$ as the encoder, and $\pxz$ as the decoder.

\subsection{VAEs and the Division of Labor}
\label{sec:divlabour}
When designing a latent variable model for image data, the decoder can be made autoregressive, where the distribution of each pixel is conditioned on the value of previous pixels in row-major order, as well as on $\z$,
\begin{equation}
    \pxz=
    \prod^N_{i=1} \prod^M_{j=1}
    p_\theta(x_{i,j}|\z, x_{<(i,j)}).
    \label{eq:full-auto}
\end{equation}
We refer to $x_{<(i,j)}$ as the full AR \emph{receptive field}, and its shape is shown in \cref{fig:explainer}b.
This modification is intended to make the model more expressive.
However, the decoder is now powerful enough to model the entire data distribution locally, to the extent that it is unclear which aspects of the data should be encoded in $\z$ and which should be modeled by the decoder.

When using a VAE in practice, there seems to be a preference to  model as much content as possible with the decoder.
He~\etal~\cite{he2018lagging} argued that this behavior is caused by the approximate posterior lagging behind the true posterior in the early stages of training, causing the parameters to get stuck in a local optimum where the decoder models the data without using the latent variables.

Alternatively, Chen~\etal~\cite{chenvariational} reasoned that with most practical VAEs, ignoring the latent variables leads a tighter lower bound on the marginal log-likelihood.
By only using the latent variables to express what the decoder cannot model, the true posterior is brought closer to the prior and can be more closely matched by the relatively inflexible approximate posterior.
The authors then demonstrated how this behavior can be used to control the division of labor in a VAE.
Specifically, they designed an AR decoder that is capable of modeling information that they do not want captured in the latent variables, but is incapable of modeling the information that they do want captured in the latent variables.

In this paper, we design an AR receptive field that only captures the correlations common in scientific imaging noise.
This allows us to train a latent variable model of noisy image data where latent variables explain the signal content and the decoder models only the noise generation process.
We then use the approximate posterior to sample latent variables, each representing one of the clean signals that could possibly underlie a given noisy image.

%% file: sections/method.tex
\section{Method}
\label{sec:method}
\figExplainer
We propose a VAE-based unsupervised image denoiser for noise that is both signal-dependent and correlated along rows or columns of pixels.
It is trained using only noisy images and does not require a pre-trained noise model.
Moreover, it does not require pixel blind-spots or subsampling.
Instead, we restrict the receptive field of the AR decoder in a hierarchical VAE~\cite{sonderby2016ladder} to a row or column of pixels, allowing the AR decoder to model the correlations of noise content but not the correlations of underlying signal.
Following the insights of Chen~\etal~\cite{chenvariational} (\cref{sec:divlabour}), the AR decoder will therefore learn to model noise, leaving the underlying signal to be encoded in the latent variable $\z$.
We then propose a method for taking these latent variables and mapping them back into image space, obtaining estimates of clean signals.
An outline of our method can be found in \cref{fig:explainer}.

\noindent\textbf{Autoregressive Receptive Fields for Noise}\quad
Our AR decoder has a one-dimensional receptive field that is sufficient for modeling row- or column-correlated noise (\cref{eq:structured}) as it occurs in microscopy (see \cref{sec:noise} \textbf{Row-Correlated Noise}) by spanning pixels in the same row or column as $x_{i,j}$.
See \cref{fig:explainer}b for a visual representation.

To remove row-correlated noise, the first step in our denoising process is to train a VAE to model noisy image data with the objective function in \cref{eq:VAEloss}, where $\pxz$ is factorized as,
\vspace{-1mm}
\begin{equation}
    \pxz 
    = \prod_{i=1}^{N} \prod_{j=1}^{M} 
    p_\theta(x_{i,j}|\z, x_{i,1},\dots,x_{i,j-1}).
    \label{eq:row_decoder}
\vspace{-1mm}
\end{equation}
To remove column-correlated noise, the factorization over pixels is perpendicular.

We find that this factorization is insufficient for modeling signal content, which is highly correlated in all directions.
Consequently, the VAE learns to encode signal content in its latent variables.
An experimental investigation of the effects of changing receptive field size can be found in \cref{sec:rfablation}, and details on how we construct this receptive field with our AR decoder architecture can be found in the supplementary material.

\noindent\textbf{Decoding the Signal}\quad
Once our VAE has been trained, its latent space will represent clean signals, \ie, each $\z$ contains all the information about an $\s$.
We would now like to use the model for denoising by inferring possible clean signals $\s$ for a given noisy image $\x$.
Unfortunately, unlike previous methods~\cite{prakash2020fully,prakashinterpretable,salmon2022towards}, we cannot directly sample clean images from our encoder.
Rather, samples from $\qzx$ will directly \emph{correspond} to a clean signal $\s$.
We denote the signal corresponding to a value of $\z$ as $\s(\z)$.
For an experimental validation of this deterministic relationship, please refer to the supplementary material.
To obtain denoised images, we approximate $\s(\z)$ with an additional regression network, termed \emph{signal decoder}, $f_\nu(\z) \approx \s(\z)$.
In the following, we describe how $f_\nu(\z)$ is trained despite not having access to training pairs $(\z,\s(\z))$.

In fact, training the signal decoder only requires pairs $(\z_k,\x_k)$, where $\x_k$ is a noisy image and $\z_k\sim q_\phi(\z|\x_k)$.
If we assume that the approximate posterior is accurate, such that $\qzx\approx\pzx$, these pairs can be equivalently thought of as samples from the joint distribution, $(\z_k,\x_k)\sim p_\theta(\z,\x)$.

Least squares regression analysis tells us that optimizing the signal decoder by minimizing the squared $L2$ norm,
\begin{equation}
\label{eq:sdloss}
    \mathcal{L}(\nu)=\lVert f_\nu(\z_k)-\x_k\rVert_2^2,
\end{equation}
will train it to approximate $\mathbb{E}_{p_\theta(\x|\z)}[\x]$ for any $\z$~\cite{hastie2009elements}.
By definition, $\z$ contains no more or less information about $\x$ than $\s(\z)$, so the signal decoder will equivalently be approximating $\mathbb{E}_{p_\theta(\x|\s(\z))}[\x]$.
Recalling that imaging noise is zero-centered (\cref{sec:noise}), the expected value of a noisy image given an underlying signal is \emph{that signal}.
Therefore, $f_\nu(\z)\approx\mathbb{E}_{p_\theta(\x|\s(\z))}[\x]=\s(\z)$.
This is similar to Noise2Noise~\cite{lehtinen2018noise2noise}, where the regressor for $\s$ is trained using noisy targets $\x$.

Even though the signal decoder would naturally be trained in a second stage after the main VAE is finished, in practice we co-trained it alongside the main VAE.
At every training step, the sampled latent variable is fed to both decoders, but only the loss from the AR decoder is allowed to backpropagate to the encoder.
This method of training is simply for convenience and we did not observe any changes in performance compared to a signal decoder that is trained separately, after the main VAE.

\noindent\textbf{Inference}\quad
\label{sec:inference}
With the VAE and signal decoder trained, we can denoise an image $\x$ in a two-step process.
We first sample a latent variable $\z \sim \qzx$, then obtain a clean image by decoding $\hat{\s}=f_\nu(\z)$.
Similarly to \cite{prakash2020fully,prakashinterpretable,salmon2022towards}, the result constitutes a random possible solution $\hat{\s} \sim p(\s|\x)$.
To obtain a consensus solution, we follow \cite{prakash2020fully,prakashinterpretable,salmon2022towards} in averaging a large number of such samples.
In the following experiments, all the results are the mean of 100 samples, both for our method and the baseline \hdn.
Please refer to the supplementary material for the inference time and PSNR of the mean of 1, 10 and 1000 samples.

%% file: sections/experiments.tex
\figVisResults
\section{Experiments}
\tabpsnr
\label{sec:experiments}
\subsection{Benchmarking Denoising Performance}
\noindent\textbf{Datasets}\quad
We tested the performance of our proposed denoiser on real noisy images captured by five different imaging modalities that commonly suffer from row-correlated noise.
Two modalities have noisy image datasets with known ground truth for quantitative evaluation.
The method for obtaining ground truth can be found in the source publications.
The first is the EMCCD sensor, for which we have three spinning disk confocal fluorescence microscopy datasets: \emph{Convallaria A}~\cite{broaddus2020removing}, \emph{Convallaria B}~\cite{prakash2020pn2v} and \emph{Mouse Actin}~\cite{prakash2020pn2v}.
The second is LSCM, for which we have three fluorescence datasets: \emph{Mouse Nuclei}~\cite{prakash2020pn2v}, \emph{Actin Confocal}~\cite{hagen2021fluorescence} and \emph{Mito Confocal}~\cite{hagen2021fluorescence}.
Note that \emph{Mouse Nuclei} contains spatially uncorrelated noise, but was included to demonstrate that the proposed method is still applicable to spatially uncorrelated noise without modification.
For details on the size of the datasets and the train/test splits, see original publications.
The remaining modalities have only noisy images.
These are the sCMOS sensor for which we have one fluorescence dataset: \emph{Embryo}~\cite{glaser2022hybrid}, the microbolometer with one infrared imaging dataset: \emph{IR}~\cite{portmann2014people},
and a STEM dataset: \emph{STEM}~\cite{henninen2020structure}.

In addition to real data, we created two datasets by corrupting the Flickr Faces HQ thumbnails dataset~\cite{karras2019style} with simulated noise.
For \emph{FFHQ - Stripes}, the images were corrupted by a combination of additive white Gaussian noise, Poisson shot noise, and additive white Gaussian noise that had undergone a horizontal Gaussian blur.
For \emph{FFHQ - Checkerboard}, the images were corrupted by a combination of a vertical checkerboard pattern and Gaussian noise with an inverse signal dependence.
Details on the simulated noises are in the supplementary materials.

\noindent\textbf{Baselines and Architecture}\quad
We compare our method with other deep learning-based denoisers that can be applied to spatially correlated, signal-dependent noise and do not require paired images.
These are the self-supervised denoisers Structured Noise2Void (SN2V)~\cite{broaddus2020removing}, Noise2Void2 (N2V2)~\cite{hock2022n2v2} and Asymmetric PD BSN (AP-BSN)~\cite{lee2022ap}, and the unsupervised denoiser Hierarchical DivNoising36 (\hdn)~\cite{prakashinterpretable}.
We use these as baseline methods along with Content-Aware Image Restoration (CARE)~\cite{weigert2018content}, which is trained using pairs of noisy and clean images, and Noise2Noise (N2N)~\cite{lehtinen2018noise2noise}, which is trained using two noisy acquisitions of the same signal, on datasets with suitable training images available.
For details on how each baseline was implemented, refer to the supplementary material.

Of the denoisers not requiring paired images, the performance of \hdn\ is the best, but the model implemented in the baseline's publication uses significantly fewer parameters than ours (7 million to 25 million).
We therefore evaluate an additional version, termed \hdn~\textit{large}, with a similar number of parameters made by increasing the number of latent dimensions from 32 to 64.

It should be noted that the \emph{Convallaria B} and \emph{Mouse Actin} datasets had been treated as spatially uncorrelated by Prakash~\etal~\cite{prakashinterpretable} when HDN was tested.
We also report those results in \cref{tabpsnr}.
It should also be noted that the noise in the \emph{Mouse Nuclei} dataset is spatially uncorrelated, so was denoised by HDN and its higher parameter version, HDN~\textit{large}, which was made the same way as \hdn~\textit{large}.

Our method requires a choice of orientation and size for the AR decoder's receptive field.
Orientation was determined by examining the spatial autocorrelation of noise samples, with these plots reported in \cref{fig:visResults}, and following the ablation study in \cref{sec:rfablation}, we always used a receptive field length of 40 pixels.

As for our model's encoder, latent variables are produced by a hierarchical VAE~\cite{sonderby2016ladder} with 14 levels to its hierarchy.
In addition to the full-sized version, we evaluate a smaller version that requires approximately 6GB (as opposed to 20GB) of GPU memory to train.
This was achieved by reducing the number of latent variables from 14 to 6 and reducing the number of latent dimensions from 64 to 32.
We refer to these models as Ours~\textit{small} and Ours~\textit{large}, for the lower memory and higher memory versions respectively.
See the supplementary materials for full architecture and training details.

\noindent\textbf{Discussion}\quad
Quantitative results measuring the Peak Signal-to-Noise Ratio (PSNR) in dB are reported in \cref{tabpsnr}, and qualitative results are reported in \cref{fig:visResults}. 
Note that \hdn\ failed to train with the \emph{Mouse Actin} and the \emph{Embryo} dataset.
Out of the methods that do not require paired images, Ours~\textit{large} achieved the highest PSNR across all datasets, even beating the supervised CARE and N2N on four of six microscopy datasets.
Ours~\textit{small} then had the second highest PSNR for an unpaired method on all datasets except \textit{FFHQ - Stripe}
.

Turning to the qualitative results in \cref{fig:visResults}, we see that Ours~\textit{large} denoised images from each dataset without leaving behind any artifacts, whereas \hdn~\textit{large} could not remove the correlated component of the noise from the \emph{STEM} or the \emph{FFHQ - Checkerboard} dataset. 
SN2V left artifacts in \emph{Convallaria A} and \emph{FFHQ - Checkerboard}.
It is unclear if the high frequency features that are visible in the \emph{Mito Confocal} output of \hdn~\textit{large} and N2V2, but not in our output, are true signal or artifacts of remaining noise, as the ground truth also contains a low level of noise.

\subsection{Ablation Study - Receptive Field Size} 
\label{sec:rfablation}
\figRFsize 
As stated in \cref{sec:method}, to model the noise correlations addressed in this paper, the receptive field of a VAE's AR decoder must span pixels in the same row or column as the pixel being predicted.
In \cref{fig:rfSize}, the effect of receptive field size (number of pixels) is investigated by denoising the \emph{FFHQ - Checkerboard} dataset using a range of receptive field lengths, from 10 to 120 pixels, and measuring the effect on PSNR.
We also include the effect of training a model with a full AR receptive field, as shown in \cref{fig:explainer}b.

This study shows that the AR decoder is able to model this spatially correlated noise, and therefore have it removed by the encoder, when the receptive field spans 40 pixels.
Moreover, the AR decoder does not seem to model more of the signal as the receptive field grows.
If it did, we would expect a steady drop in PSNR as denoised images lose signal content.
However, as the image on the right of \cref{fig:rfSize} shows, the signal will be modeled by an AR decoder with a full receptive field.
Latent variables in this situation carry no information about $\x$, causing the signal decoder to minimize \cref{eq:sdloss} by predicting the mean of the training set.
This is expected, as when the latent variable $\z$ has no information about $\x$, $\mathbb{E}_{p_\theta(\x|\z)}[\x]=\mathbb{E}_{p_\theta(\x)}[\x]$.

\subsection{Ablation Study - Noise Reconstruction}
\figNoiseRecon
If the VAE's AR decoder is modeling only the noise component of images, encoding an image with the VAE and sampling from the AR decoder should yield an image with the same underlying signal but a different random sample of noise.
If the decoder's model of the noise is accurate, the sampled noise should exhibit the same autocorrelation and signal-dependence characteristics as the noise in the original image.
An investigation into this is reported in \cref{fig:noiserecon}, where a noisy image from the \textit{Convallaria A} dataset was encoded by a trained VAE, then a latent variable was sampled and used by the VAE's AR decoder to sample a reconstruction of the original image.
The reconstructed noise exhibits spatial autocorrelation and signal dependence very similar to the real noise, indicating that our AR decoder has learnt an accurate model of the noise.

%% file: sections/conclusion.tex
\vspace{-2mm}
\section{Conclusion}
\label{sec:conclusion}
\vspace{-1mm}
We proposed an unsupervised VAE-based denoising algorithm for signal-dependent noise that is correlated along rows or columns of pixels.
It is trained without any clean data or a noise model.
By engineering the receptive field of the AR decoder, the VAE's latent variables are encouraged to represent the signal content of an image while discarding the noise.
We then presented a novel signal decoder that is trained to map this latent variable to an estimate of the clean image.
The algorithm outperforms existing self- and unsupervised denoisers.

Our method is suited to noise with correlations that run parallel to the image axes.
This commonly occurs in microscopy and lab users often cannot find suitable methods to remove it.
We release our code open source and strongly believe that the scientific imaging community will apply and adapt our methods in a variety of applications.

While we have achieved our results using 1-dimensional receptive fields, some imaging modalities produce noise that is correlated in multiple directions, therefore requiring a differently shaped receptive field.
We found that extending the receptive field to cover both a row and a column of pixels allows the AR decoder to model some aspects of the signal, making the technique unsuitable for removing noise correlated in two dimensions.
However, we believe that techniques other than shaping the AR decoder's receptive field could be discovered to limit it's modeling capabilities.
Furthermore, the method is limited to zero-centered noise by the signal decoder.
This precludes, \eg, removing salt-and-pepper noise or restoring images with saturated pixel intensities.
A direction for future work could therefore be alternative methods for mapping latent variables back into image space. 
We hope that future work will further improve the theoretical understanding of the method and allow us to utilize its full potential.

\noindent{\textbf{Acknowledgements}}
We thank Ale\v{s} Leonardis (University of Birmingham) for a helpful discussion on the writing of this paper.
Computations in this paper were performed using the University of Birmingham's BlueBEAR HPC service, which provides a High Performance Computing service to the University's research community. See \url{http://www.birmingham.ac.uk/bear} for more details.

%% file: supp_sections/Latent_variables_represent_clean_images.tex
\section{Latent Variables Represent Clean Images}
The training of the signal decoder assumes that every sampled value of the latent variable $\z$ corresponds one clean image, or signal, $\s$.
We denote the signal corresponding to a value of $\z$ by $\s(\z)$.
Using this relationship, the signal decoder, $f_\nu(\z)$, can be trained to estimate 
\begin{equation}
\label{eq:sz}
    f_\nu(\z)\approx\mathbb{E}_{p_\theta(\x|\z)}[\x]=\mathbb{E}_{p_\theta(\x|\s(\z))}[\x]=\s(\z).
\end{equation}
Here, we provide an another way of viewing this deterministic relationship.

If the latent variables of our model truly represent only signals, then the AR decoder, $\pxz$, must model only the noise generation process.
Therefore, different random samples from the AR decoder for the same value of latent variable will be images with the same underlying signal and different random samples of noise.
Since the noise is zero-centered, this allows us to produce an estimate of the signal by calculating the mean of many samples from the AR decoder.
That is, if $\x_1,\x_2,\ldots,\x_L$ are $L$ random samples from $\pxz$,
\begin{equation}
    \overline{\x} =
    \frac{1}{L} \sum^L_{l=1}\x_l
    \approx\mathbb{E}_{p_\theta(\x|\z)}[\x]
    =\s(\z)
    .
\end{equation}

In this section, we experimentally verify \cref{eq:sz} by estimating the signal underlying a noisy image $\x$ using both techniques; by passing a latent variable sample to the signal decoder and by averaging 10,000 noisy image samples from the AR decoder.
\figSz
If \cref{eq:sz} is true, the two estimates of the signal should be nearly identical for the same value of latent variable.

\Cref{fig:Sz} shows the result of this experiment for two different random samples from the approximate posterior, $\z_1$ and $\z_2$.
The two estimates of $\s(\z_1)$ are visually very similar to each other, while exhibiting clear structural differences from the two estimates of $\s(\z_2)$.


%% file: supp_sections/Training_and_inference.tex
\section{Training and Inference}
\subsection{Hyperparameters}
Both the main VAE and the signal decoder were trained with an Adamax~\cite{kingma2014adam} optimizer with a learning rate of 0.002.
Both learning rates decreased by a factor of 10 when the validation loss had plateaued for 50 epochs.
The models for all datasets were trained for a maximum of 80,000 steps but stopped if validation loss had plateaued for 100 epochs.

For the non-simulated datasets, training images were randomly cropped to a size of $256\times256$ at each epoch and a batch size of 16 was used, but this was split into 4 virtual batches.
For the \emph{FFHQ - Stripe} and \emph{FFHQ - Checkerboard} datasets, training images were kept at their original resolution of $128\times 128$ and a batch size of 64 was used, but split into 16 virtual batches.

\subsection{Hardware and Software}
The workstation used for this paper's experiments is a 36 core Intel(R) Core(TM) i9-10980XE CPU @ 3.00GHz with 134GB of RAM running Ubuntu 22.04.4, Python 3.11.5 and pytorch-lightning 2.2.1. The GPU used for training is a NVIDIA GeForce RTX 3090 with 24GB of VRAM. 
For all datasets, training required approximately 20GB of GPU memory with the \emph{large} network and 6GB with the \emph{small} network.

\subsection{Times}
Training a model for 80,000 steps with our hardware takes approximately 24 hours.
After training, denoising 100 images of size 512x512 by randomly sampling 100 denoised estimates takes 13 minutes when using a batch size of 10.
Increasing batch size made no improvement as the GPU is at 100\% utilization.

\subsection{PSNR of Minimum Mean Square Error Estimates}
\tabsampling
In addition to the quantitative results presented in the paper, the PSNR of the mean of 1, 10 and 1000 random denoised samples can be found in \cref{tab:sampling}.

%% file: supp_sections/Architecture.tex
\section{Architecture}
We proposed two network architectures for denoising, one \emph{large} and one \emph{small}.
Each model consists of a hierarchical Variational Autoencoder (VAE)~\cite{sonderby2016ladder}, an autoregressive decoder~\cite{van2016conditional} and our novel \emph{signal decoder}.
In the \emph{large} network, the VAE has 14 levels in its hierarchy.
The first 13 levels have 64 latent dimensions each, while the final level has 128 dimensions. 
The latent variable passed to the decoders is sampled from this final level.
At each level on both the bottom-up path and the top-down path is a residual block consisting of two sets of a convolution followed by a batch normalization~\cite{ioffe2015batch} followed by a Mish activation function~\cite{misra2019mish}.
Each residual block is followed by a gated block~\cite{vahdat2020nvae}.
Resampling is performed at alternating levels.
The \emph{small} network is the same except that it has 6 levels to its hierarchy and half the latent dimensions.

The autoregressive decoder is built with eight layers of conditional PixelCNN blocks as proposed in~\cite{van2016conditional}, but we found the performance to be better with a ReLU activation function~\cite{agarap2018deep} than with gated units.
The convolving kernels in the AR decoder have dimensions $1\times k$, where $k$ is the kernel size.
In the first layer of the decoder, the input is padded with $k$ zeros on its left-hand side, and then a convolution is applied.
At all subsequent layers, the input features are padded with $k-1$ zeroes on the left-hand side.
This results in a row-based autoregressive receptive field.
For a column-based receptive field, the kernels have dimensions $k\times 1$ and padding is applied to the top of the input.
For all of our experiments, $k=5$.
The convolutions in every other layer have dilated kernels~\cite{yu2015multi} and all have 64 filters.
The likelihood distribution is a Gaussian mixture model, with 3 components used for all datasets except the \emph{FFHQ} datasets, for which 10 were used, and the \emph{STEM} dataset, for which 5 were used.

The \emph{signal decoder} is a convolutional neural network consisting of four 3$\times$3 convolutions with 128 filters, each followed by a ReLU activation function.

%% file: supp_sections/Baselines.tex
\section{Baselines}
\noindent\textbf{AP-BSN~\cite{lee2022ap}}\quad
These models were trained using the code available at \url{https://github.com/wooseoklee4/AP-BSN} using hyperparamters detailed in the original publication.
We used a stride factor of 5 for all datasets.

\noindent\textbf{Structured Noise2Void~\cite{broaddus2020removing} and N2V2~\cite{hock2022n2v2}}\quad
Both these model types were trained using the code available at \url{https://github.com/juglab/n2v}, using default hyperparameters found in the example notebooks for SN2V and hyperparameter values found in the original publication of N2V2.
Following Broaddus~\etal~\cite{broaddus2020removing}, SN2V masks should be as small as possible while covering pixels with a noise value that is highly predictive of the noise value in the target pixel.
A trial and error test of the mask size for each dataset would be too computationally expensive, so we follow~\cite{broaddus2020removing} and mask 4 pixels on each side of the target pixel for all datasets except \emph{FFHQ - Checkerboard}.
The structured component of noise in the \emph{FFHQ - Checkerboard} dataset can theoretically be predicted by seeing only two pixels in the same column, so entire columns were masked here.
The orientation of the pixel mask was determined by looking at the spatial autocorrelation in noise patches for each dataset.
The \emph{Mouse Nuclei} dataset is corrupted by unstructured noise, so was denoised with a single pixel mask.

\noindent\textbf{\hdn~\cite{prakashinterpretable}}\quad
These models were trained using the code available at \url{https://github.com/juglab/HDN/} using default hyperparameters found in the example notebooks.
\hdn~requires a pre-trained noise model.
We followed Prakash~\etal~\cite{prakashinterpretable} and modeled the noise in each dataset using a Gaussian mixture model.
The noise model parameters can be estimated from the training data of datasets with available ground truth.
For datasets without ground truth, we trained the noise model using denoised images from our method as pseudo-ground truth.

\noindent\textbf{CARE~\cite{weigert2018content} and N2N~\cite{lehtinen2018noise2noise}}\quad
Both of these model types were trained using the code available at \url{https://github.com/CSBDeep/CSBDeep}, using default hyperparameters and setting noisy images as target for N2N.

%% file: supp_sections/Simulated_data.tex
\section{Simulated Data}
The Flickr Faces HQ thumbnails dataset~\cite{karras2019style}, with resolution $128\times128$, was made grayscale by averaging across color channels.
For \emph{FFHQ - Stripe}, the ground truth, $\s$, was scaled to have pixel values between $0$ and $1$, and Poisson noisy images were created as $\x=0.002\times \mathcal{P}(\s/0.002)$.
Zero-mean Gaussian noise with a standard deviation of $0.02$ was then added to these images.
Finally, structured noise was created by applying a horizontal Gaussian blur with a standard deviation of $1$ to white Gaussian noise with a standard deviation of $0.025$ and added on top.
For \emph{FFHQ - Checkerboard}, we added noise with inverse signal dependence by sampling Gaussian noise from the distribution $\mathcal{N}(0, 0.15\times 1/\s)$.
Then a vertical checkerboard pattern was added by subtracting $0.1$ from two pixels and adding $0.1$ to the next two pixels along columns.
The starting point for the checkerboard was randomly sampled from a uniform distribution.
For both \emph{FFHQ} datasets, the final 1000 images were designated as a test set.

%% file: supp_sections/qualitative_results.tex
\section{Additional Qualitative Results}
See overleaf for larger denoised images from each dataset.
\figConvA
\figConvB
\figMouseA
\figMouseN
\figActinConfocal
\figMitoConfocal 
\figEmbryo
\figStem 
\figInfrared
\figStripe
\figCheckerboard